# Random-phase spatial solitons in instantaneous nonlocal nonlinear media

Oren Cohen, Hrvoje Buljan, Tal Schwartz, Jason W. Fleischer, and Mordechai Segev

*Physics Department, Technion - Israel Institute of Technology, Haifa 32000, Israel*

We predict random-phase spatial solitons in instantaneous nonlocal nonlinear media. The key mechanism responsible for self-trapping of such incoherent wave-packets is played by the non-local (rather than the traditional non-instantaneous) nature of the nonlinearity. This new kind of incoherent solitons has profoundly different features than other incoherent solitons.

The observation of spatially-incoherent optical solitons [1] has opened a new direction in nonlinear science [1-7]. Such spatially-incoherent solitons – self-trapped entities whose structure varies randomly in time – form when diffraction, governed by the beam's correlation function, is robustly balanced by nonlinear self-focusing. This balance results in the stationary propagation of the beam's envelope (i.e. its time-averaged intensity) [3]. A prerequisite for the formation of spatially incoherent solitons is that the nonlinearity responds to the envelope of the beam rather than to the fluctuating intensity pattern. Otherwise, the speckled nature of the field would induce intricate spatial variations in the refractive index, causing beam fragmentation, and prohibiting self-trapping. In their original concept, incoherent solitons were studied with a non-instantaneous nonlinearity having a response time, $t$, much longer than the characteristic beam fluctuation time $t_c$ ($t >> t_c$). There, the nonlinearity time-averages over the stochastic multi-mode character of the instantaneous speckled field [3], responding only to the envelope of the beam. In fact, spatially incoherent solitons have been studied only in rather slow nonlinear media [1-5], and it has been believed that a non-instantaneous response is a prerequisite for the formation of such solitons [3,6]. Here, we show that if the nonlinearity has a *nonlocal* nature, it can filter out the otherwise highly fragmented variations in the refractive index induced by the rapidly fluctuating multimode field. In this fashion, *incoherent (random-phase) spatial solitons can form in instantaneous nonlocal nonlinear media.*

Nonlocal nonlinearities are inherent in many systems, when the underlying mechanism involves transport (of heat [8], atoms in a gas [9], etc.), or long-range forces (e.g., electrostatic interactions in liquid crystals [10]). Nonlocality also affects the propagation of waves in plasma [11], and matter waves in Bose-Einstein condensates, where nonlocality arises from the underlying many-body interactions [12]. For localized wavepackets, of which solitons are an exemplary phenomenon, nonlocality becomes important when the range of nonlocal interactions is appreciable on the scale of the wavepacket width. Nonlocality has profound consequences on solitons [13-15], by arresting catastrophic collapse [14], giving rise to attraction between dark solitons (that otherwise repel) [15], etc.

Here, we predict a fundamentally new type of incoherent soliton, forming in instantaneous nonlocal nonlinear medium ($t_c >> t$). These solitons form when (i) the beam is self-trapped within a time frame much shorter than $t_c$, and (ii) the transverse momentum of the beam is constant in time. When the latter condition is violated, the time-averaged beam exhibits a new type of propagation-broadening mechanism: *statistical nonlinear diffraction*.

The propagation of weakly-correlated waves in fast-responding nonlocal nonlinear media is an issue of generic interest. Here, we analyze this problem in the context of optics. Consider a quasi-monochromatic partially-spatially-incoherent beam, propagating in a fast-responding nonlocal nonlinear medium. The characteristic speckle size ($\approx$ transverse correlation distance) is at least several times larger than the wavelength; hence the paraxial approximation is valid. The characteristic time scales involved are $t$, $t_c$, and the "time of flight" $t_f$ during which a light beam passes through the medium. Here, we consider the regime in which $t \ll t_c$ and $t_f \ll t_c$. For the sources typically used to study random-phase solitons (e.g. rotating diffuser [1]), the coherence time is a controllable variable. Hence, the relations above and the nonlinear dynamics we suggest below are experimentally accessible.

The complex field $\Psi(x,z,t)$ describing the spatially-incoherent light is fluctuating with a characteristic time-scale $t_c$. We study the propagation of $\Psi(x,z,t)$ in two steps. First, we analyze the propagation within a very short time interval ($\ll t_c$) during which the beam can be treated as a **coherent** multimode (speckled) wave. Second, we calculate the propagation of the time-averaged envelope [16]. Assuming the incoherent light source is ergodic, the time average corresponds to an ensemble average over all possible realizations of the speckled field. In addition, because the response time of the nonlinear medium is very short, the nonlinearity has no memory, and we resort to ensemble averaging while analyzing the propagation of the (time-averaged) beam envelope.

First we study the dynamics within the short time frame. The dimensionless slowly-varying amplitude of the optical field $\Psi(x,z,t)$ evolves according to

$$i\frac{\partial \Psi}{\partial z} + \frac{\partial^2 \Psi}{\partial x^2} + \Delta n[|\Psi|^2]\Psi(x,z,t) = 0, \qquad (1)$$

where *x* and *z* are the transverse and propagation directions, respectively. The nonlocal nonlinear term $\Delta n$ has the form of a spatial convolution between the instantaneous wave intensity $|\Psi(x,z,t)|^2$ and the response function R(x,x') = R(x-x') of the medium [14]

$$\Delta n(x,z,t) = \int_{-\infty}^{\infty} R(x'-x)|\Psi(x',z,t)|^2 dx'. \qquad (2)$$

For concreteness, consider a Gaussian response function $R = \frac{1}{\sqrt{\piس^2}} \exp[-(x-x')^2/س^2]$. Here, we are interested in the highly nonlocal regime, occurring when the width of the response function $س$ is much larger than the width of the beam. In this regime, the nonlinear index change, to a good approximation, averages over the variations in the beam intensity profile, and $\Delta n$ has a parabolic shape, depending only on the total power [13,14]

$$\Delta n(x,z,t) \approx \frac{P(t)}{\sqrt{\pi س^2}} \left\{ 1 - \frac{[x-a(z,t)]^2}{س^2} \right\}, \qquad (3)$$

where $P(t) = \int_{-\infty}^{\infty} |\Psi(x,z,t)|^2 dx$ is the total power of the beam within this time frame, and $a(z,t) = \int x|\Psi(x,z,t)|^2 dx$ is the "beam center" at plane $z$. The center of the induced waveguide coincides with the center of the beam $a(z,t)$. The beam enters the nonlinear medium at angle $\theta(t)$ (with respect to z) which varies stochastically with time. This propagation angle corresponds to the initial transverse momentum of the beam: $\theta(t) = 2\bar{k}_x(t) = 2\int_{-\infty}^{\infty} k_x|\tilde{\Psi}(k_x,z=0,t)|^2 dk_x$ (the factor 2 is because in our dimensionless units, $k_z=1/2$). Because Eq. (1) conserves transverse momentum, the angle does not change along z, and the center of the waveguide $a(z,t)$ lies on a straight line: $a(z,t) = a(0,t) + 2\bar{k}_x(t)z$.

The discussion above assumes that the width of the beam is much smaller than the nonlocality range $س$ for all z. Let us examine when this happens. Consider a beam $\Psi(x,z=0,t)$ that at $z=0$ is much narrower than the nonlocality range $س$. In this limit, the beam induces a parabolic waveguide at the vicinity of the input face, and some of its guided modes are excited by $\Psi(x,z=0,t)$. If all the modes excited by the beam are much narrower than $س$, the highly nonlocal limit is satisfied throughout propagation. In this situation, the instantaneous induced waveguide is stationary with a straight line trajectory of a(z,t), while the beam is populating its guided modes in a self-consistent fashion [17].

**The instantaneous beam is thus self-trapped,** yet its intensity oscillates periodically due to "beating" among the modes comprising it.

Interestingly, having a self-trapped speckled beam at any instantaneous time frame does not necessarily guarantee that the time-averaged behavior of such a beam exhibits self-trapping. This is because the initial propagation angle of the beam, $q(t)$, and its transverse displacement $a(0,t)$, fluctuate randomly on time-scale $t_c$. We therefore examine the time-average behavior of the system, with the average taken over $t \gg t_c$. From ergodicity, such averaging is equal to an ensemble average over all possible initial conditions $q(t)$ and $a(0,t)$. Let us denote $P(\bar{k}_x)$ as the probability distribution of the transverse momentum of the incident beam. Consider first the case where $\bar{k}_x(t)$ is a random variable, hence $P(\bar{k}_x)$ has some width. When self-consistency is satisfied (the beam is self-trapped in each frame), the instantaneous intensity at a large enough z is located in the vicinity of the beam center, $a(z,t) \approx 2\bar{k}_x(t)z$. Thus, the time-averaged intensity $\langle |\psi(x,z,t)|^2 \rangle$ after distance z is proportional to $P(\bar{k}_x)z$. That is, the time-averaged intensity *broadens*, with a width proportional to z and to the probability distribution of the transverse momentum of the light (defined by the source), $P(\bar{k}_x)$. *Consequently, an incoherent beam in a nonlocal nonlinear medium may form self-trapped solitonic beams in each short time-frame, while its time-averaged intensity structure is broadening.* We emphasize that this propagation-broadening is nonlinear, arising because an incoherent source typically emits light with stochastically-varying directionality. Henceforth we address this new propagation-broadening mechanism as *statistical nonlinear diffraction*.

There are cases, however, when the statistical nonlinear diffraction is eliminated. Such cases occur, for example, when the source emitting the incoherent light does not have randomly fluctuating transverse momentum, i.e., when $P(\bar{k}_x) = d(\bar{k}_x)$, thus forming instantaneous self-trapped beam with $q = 0$ at all times t. In this case, the **beam self-traps within each short time frame and also form a time-averaged random-phase soliton**.

Let us now analyze some examples. Consider first a beam which at z=0 is a superposition of two uncorrelated coherent Gauss-Hermite waves

$$\Psi(x,t,z=0) = \frac{\exp(-x^2)}{\sqrt[4]{2\pi}} \left[ \sqrt{\frac{P_n}{2^n n!}} H_n(\sqrt{2}x) + \sqrt{\frac{P_m}{2^m m!}} H_m(\sqrt{2}x) \exp(i\varphi(t)) \right] \quad (4)$$

where $n \neq m$, $H_n$ is the Hermite polynomial of order n, $P_n$ is the modal power of wave n, and $\varphi(t)$ is a real random variable uniformly distributed in the interval $\{-\pi, \pi\}$. Such a beam, with $P_n$ and $P_m$ being constants and $\varphi(t)$ stochastic, has been used in the past to generate multimode solitons [18]. In the highly nonlocal limit, this beam induces a parabolic waveguide whose width depends only on the total power $P = P_n + P_m$. First, we set $P_n = P_m = 2\sqrt{\pi}\sigma^3$ ($P = 4\sqrt{\pi}\sigma^3$). For such P value, the uncorrelated waves of $\Psi(x, z=0, t)$ coincide with the guided (Gauss-Hermite) modes of the induced waveguide. To work out the time-averaged propagation of such a beam in our system, we average over the propagation of 100 of its realizations ($\varphi_j = -\pi + 2\pi j/100 \quad j = 1,2,3...100$). In each frame, we simulate the evolution of the coherent beam, $\Psi_i$, [via Eq. (1)], using a standard BPM code. In the simulations, we do not approximate the waveguide by Eq. (3), but calculate the convolution integral [Eq. (2)] assuming a Gaussian response function with $\sigma = 20$.

Figure 1 presents the results with n=0 and m=1. Figures 1a and 1b show two representative frames: $\varphi = 0$ (Fig. 1a) and $\varphi = \pi/2$ (Fig. 1b). Within each frame, the beam is self-trapped, yet the instantaneous beams are propagating with a fast fluctuating directionality. Consequently, the time- (ensemble-) averaged beam (Fig. 1c) broadens due to statistical nonlinear diffraction. For comparison, we simulate the linear propagation of the time-averaged intensity (Fig. 1d). After some distance, (shown in Fig. 1e for z=3) the statistical nonlinear diffraction (dashed) leads to a different beam profile compared to the profile of the linearly diffracting beam (solid), although the two intensity maxima in both cases coincide. The spatial power spectrum and the probability distribution of the transverse momentum are plotted in Fig. 1f. The calculated profiles (Fig. 1f) resemble the calculated linear and nonlinear diffraction profiles (Fig. 1e). This shows that the far-field time-averaged intensity structure of the statistical nonlinear diffraction indeed corresponds to the probability distribution of the transverse momentum of the incident beam.

An incoherent soliton forms in our system if its statistical nonlinear diffraction is eliminated, i.e. the transverse momentum is constant (not stochastic). It is simple to show that for $n \neq m \pm 1$ the transverse momentum of $\Psi(x, z=0, t)$ is always zero, i.e. $P(\bar{k}_x) = \delta(\bar{k}_x)$. Figure 2 shows an example of a soliton comprised of modes 0 and 2. Figures 2a and 2b show two representative frames: $\varphi = 0$ (Fig. 2a) and $\varphi = \pi/2$ (Fig. 2b). Now, in every time frame, the beam not only self-traps, but is also always propagating exactly on axis. The ensemble-average is shown in Fig. 2c, demonstrating the stationary propagation of the time-averaged envelope. Figure 2c is a representative multimode soliton occurring when the uncorrelated coherent waves of the incident light coincide with the guided modes of the induced waveguide. When the guided modes do not exactly coincide with the uncorrelated coherent waves of the incident light, the waveguide modes are excited with some correlation among them. Consequently, "beating" among the (partially-correlated) guided modes will now lead to oscillations of the envelope along propagation. If statistical nonlinear diffraction is eliminated, this time-averaged beam forms a "soliton-breather". Such an example is shown in Fig. 2d, where the modal powers of the incident beam are 20% smaller than those of Fig. 2c. This results in a 20% shallower induced waveguide, supporting guided modes with a different structure than the structure of the incident beam. Next, we study the case in which the modal powers fluctuate randomly. In this case we calculate the propagation of the time-averaged beam via Monte-Carlo method. Figure 2e shows the propagation of a beam in which each modal power has a Gaussian distribution with an average $\langle P_n \rangle = \langle P_m \rangle = 2\sqrt{\pi}\sigma^3$ (values of Fig. 2c) and standard deviation $\sigma_n = 0.2 \langle P_n \rangle$. As shown there, the beam self-traps into an incoherent "soliton-breather". For comparison, Fig. 2f shows the linear diffraction of the time-averaged beams.

As a last example, consider a quasi-thermal light beam (e.g. a laser beam has passed through a rotating diffuser [1]) propagating in our system. Obviously, quasi-thermal light excites consecutive modes of the induced waveguide, and hence statistical nonlinear diffraction is always present. However, for highly incoherent beams, the number of excited modes becomes very large and hence the ratio of consecutive pairs to inconsecutive pairs can be very small. In this case the contribution of statistical nonlinear diffraction to the propagation of the time-averaged beam is very small. For example, consider a beam consisting of 50 coherent waves $\Psi(x,t,z=0) = \frac{\exp(-x^2)}{\sqrt[4]{2\pi}} \sum_{n=0}^{n=49} \sqrt{\frac{P_n(t)}{2^n n!}} H_n(\sqrt{2}x)\exp(i\varphi_n(t))$ incident upon the nonlinear medium. $\varphi_n$ are statistically independent random variables. The modal powers, $P_n(t)$, are random variables having a Gaussian distribution with $\langle P_n \rangle = P\exp(-n/\Delta) \Big/ \sum_{n=0}^{n=49} \exp(-n/\Delta)$, $\Delta = 25$, $P = 4\sqrt{\pi}\sigma^3$ and $\sigma_n = 0.2\langle P_n \rangle$. Figures 3a and 3b show the propagation of two realizations of the incident beam. In each frame the beam is self-trapped, yet the beam in different frames is propagating in a different direction, producing the statistical nonlinear diffraction of the time-averaged envelope [Fig. 3c]. However, since the beam has 50 modes, the statistical nonlinear diffraction is very small compared to the linear diffraction [Fig. 3d; note the 4.5 times difference in the transverse scales between Figs. 3(a-c) and Fig. 3d]. Figure 3e shows the sharp contrast between the linear (solid) and the nonlinear (dashed) diffraction effects for the time-averaged normalized intensity. Clearly, the broadening is very large in the linear case, whereas the nonlinear broadening is practically negligible for the highly multimode beam. Recalling that for the bi-modal beam the statistical nonlinear diffraction is comparable to the linear diffraction (fig. 1e), implies that the ratio between the linear and nonlinear broadening increases as the number of modes increase. That is, for the same value of nonlinearity, the more incoherent the beam, the more stationary its time-averaged intensity. There is no "penalty" for making the self-trapped beam more incoherent in a highly nonlocal nonlinear medium; in fact, lower spatial coherence results in "better" self-trapping (as long as self-consistency is satisfied). Finally, the power spectrum and the probability distribution of the transverse momentum highlight the similarity between the spectral profiles (Fig. 3f) and the calculated diffraction profiles (Fig. 3e).

In conclusion, we have studied the propagation of spatially-incoherent beams in a fast responding (instantaneous) nonlocal nonlinear medium. A soliton can form in this system when (i) the beam is self-trapped within a time frame much shorter than $t_c$, and (ii) the transverse momentum of the incident beam is constant in time. When the transverse momentum randomly fluctuates in time, the beam exhibits a new kind of diffraction-broadening denoted as statistical nonlinear diffraction. The theory presented was analyzed with parameters readily accessible to experiment, and it is anticipated that nonlinear dynamics of such nonlocal spatially-incoherent beams will be observed in the near future.

**Figure captions**

Figure 1: Propagation of a random-phase beam comprised of Gauss-Hermite waves 0 and 1. Single frame self-trapped propagation when (a) $\varphi = 0$, and (b) $\varphi = \pi/2$. Inserts show the induced waveguides. Propagation of time-averaged intensity with nonlinearity "on" demonstrating statistical nonlinear diffraction (c), and with nonlinearity "off" showing linear diffraction (d). Note that the x-scale in (d) is 2.5 times larger than in (a-c). (e) Normalized time-averaged intensity profiles at the input face (dotted), and at the output face with (dashed) and without (solid) nonlinearity. (f) Calculated power spectrum of the beam (solid) and probability distribution of the transverse momentum (dashed).

Figure 2: Propagation of random-phase beams consisting of Gauss-Hermite waves 0 and 2. Single frame beam propagation when (a) $\varphi = 0$, and (b) $\varphi = \pi/2$. Propagation of the time-averaged intensity when the incident waves coincide (c) [do not coincide (d)] with the guided modes of the waveguide demonstrating a soliton-breather. (e) Propagation of the time-averaged intensity of an incoherent beam demonstrating an incoherent soliton. (f) Propagation of the time-averaged intensity with nonlinearity "off" showing linear diffraction. The x-scale in (f) is 5 times larger than in (a-e).

Figure 3: Propagation of an incoherent beam from a quasi-thermal source. (a,b) propagation of individual frames. Propagation of the time-averaged intensity with nonlinearity "on" demonstrating self-trapping with a small statistical nonlinear diffraction (c), and with nonlinearity "off" showing linear diffraction (d). The x-

scale in (d) is 4.5 times larger than in (a-c). (e) Normalized profiles of the time-averaged intensity at the input face (dotted), and output face with (dashed) and w/o (dashed) the nonlinearity. (f) Time-averaged power spectrum of the beam (solid) probability distribution of the transverse momentum (dashed).

Figure 1

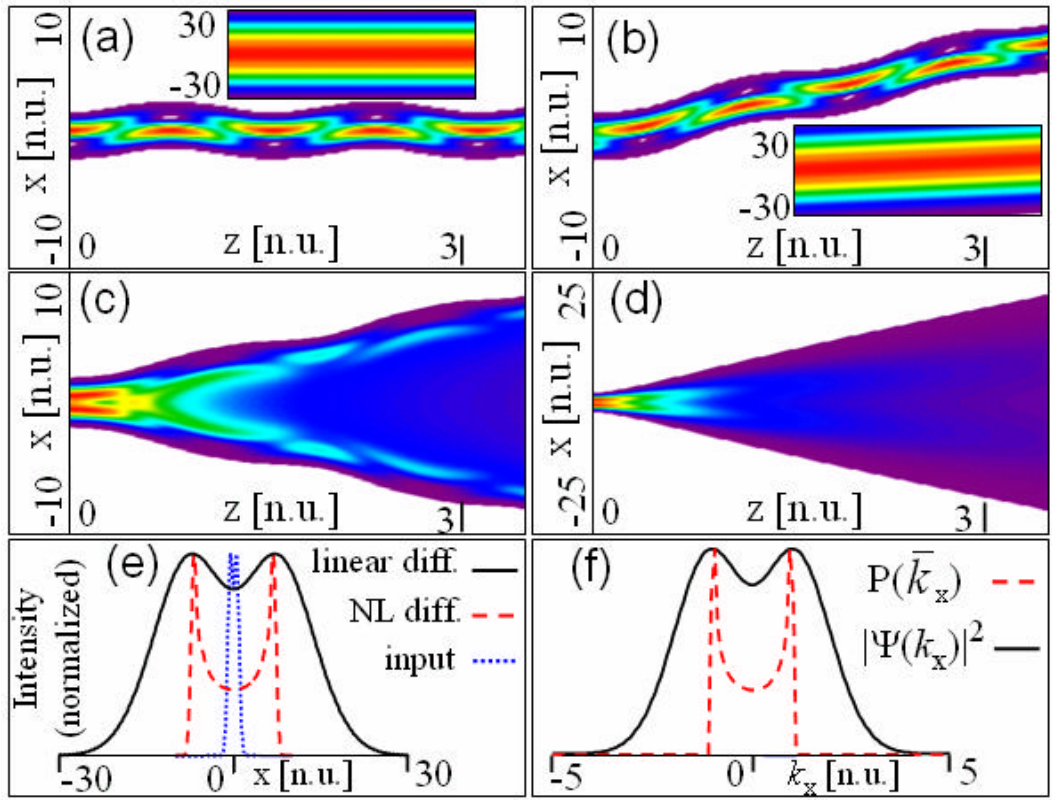

Figure 2

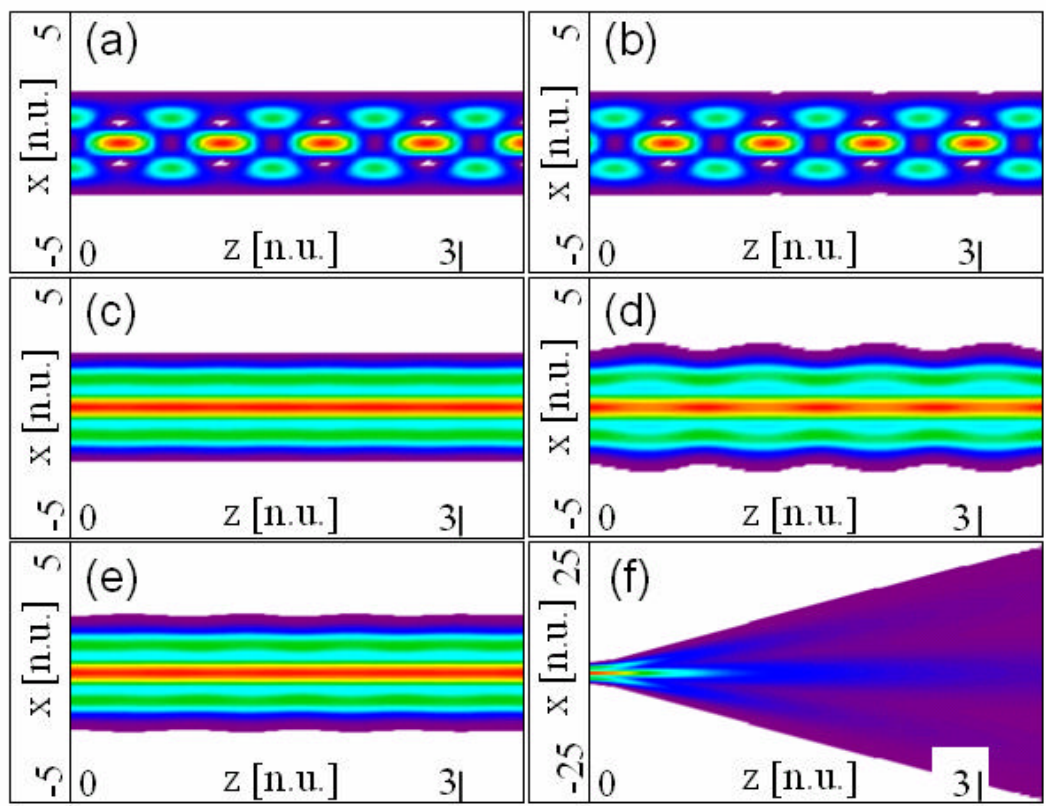

Figure 3

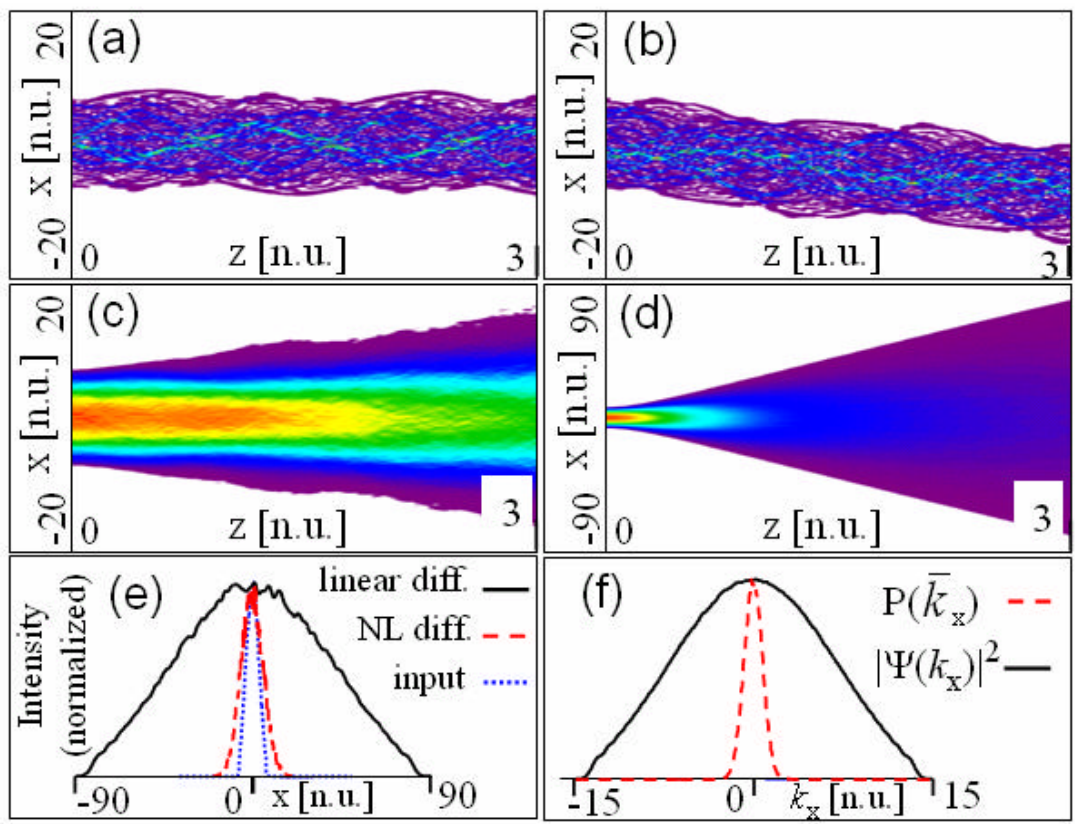